\documentclass[11pt,a4paper]{article}

\usepackage{graphicx}
\usepackage{epsfig}
\usepackage[titletoc,toc,title]{appendix}
\usepackage[dvips]{color}

\usepackage{hyperref}
\usepackage{graphicx,epsf}
\usepackage{bm}
\usepackage{amsmath}

\begin{document}

\begin{center}
{\Large{\textbf{Decay of geodesic acoustic modes due to the combined action of phase mixing and Landau damping.}}}\\
\vspace{0.2 cm}
{\normalsize {A. Biancalani$^{1}$, F. Palermo$^{1}$, C. Angioni$^{1}$, A. Bottino$^{1}$, F. Zonca$^{2,3}$.\\}}
\vspace{0.2 cm}
\small{${}^1$ Max-Planck Institute for Plasma Physics, 85748 Garching, Germany \\
${}^2$ ENEA C. R. Frascati, Via E. Fermi 45, CP 65-00044 Frascati, Italy\\
${}^3$ Inst. for Fusion Theory and Simulation, Zhejiang University, 310027 Hangzhou, P. R. China\\
\vspace{0.2cm}
contact of main author: www2.ipp.mpg.de/$\sim$biancala}
\end{center}

\begin{abstract}
Geodesic acoustic modes (GAMs) are oscillations of the electric field whose importance in tokamak plasmas is due to their role in the regulation of turbulence. The linear collisionless damping of GAMs is investigated here by means of analytical theory and numerical simulations with the global gyrokinetic particle-in-cell code ORB5. The combined effect of the phase mixing and Landau damping is found to quickly redistribute the GAM energy in phase-space, due to the synergy of the finite orbit width of the passing ions and the cascade in wave number given by the phase mixing.
When plasma parameters characteristic of realistic tokamak profiles are considered, the GAM decay time is found to be an order of magnitude lower than the decay due to the Landau damping alone, and in some cases of the same order of magnitude of the characteristic GAM drive time due to the nonlinear interaction with an ITG mode. In particular, the radial mode structure evolution in time is investigated here and reproduced quantitatively by means of a dedicated initial value code and diagnostics.
\end{abstract}

\section{Introduction}

Plasmas in magnetic-confinement-fusion plasmas are turbulent, due to the presence of strong spatial gradients. Turbulence is often observed accompanied by zonal (i.e. axisymmetric) radial electric fields, generating poloidal ExB flows in tokamaks. These zonal flows are usually constituted by two components, one nearly constant in time, dubbed zero-frequency zonal flows (ZFZFs)~\cite{Hasegawa79,Rosenbluth98,Diamond05,Palermo15pop,Palermo15epj,Ghizzo15popa,Ghizzo15popb}, and one oscillating with a frequency of the order of the sound frequency ($\omega \sim c_s /R$, where $c_s$ is the sound speed and $R$ is the tokamak major radius), dubbed Geodesic Acoustic Modes (GAMs)~\cite{Winsor68,Zonca08}.
The importance of both components is due to their nonlinear interaction with turbulence, for they shear and distort convective and turbulent cells leading to the turbulence saturation and consequently to a reduction of the transport.
The peculiarity of GAM oscillations resides in the different shearing efficiency that the zonal flows have in relation to their oscillatory behavior. In fact, while ZFZFs suppress turbulence efficiently, oscillations make the action of the zonal 
flow less effective~\cite{Miyato04,Angelino06}. ZFs/GAMs are also known to transfer energy to the long-wavelength components of the turbulence~\cite{Scott92,Scott05}.


Historically, the Landau damping has been recognized as one of the main damping mechanisms of GAMs. The Landau damping is due to the nonuniformity of the ion distribution function in velocity-space, and it is therefore an intrinsically kinetic effect.
Its evaluation has been performed with increasing accuracy, where finite-orbit-width (FOW) of the passing ions were initially neglected~\cite{Zonca96,Zonca08}, then included to the first order~\cite{Zonca98,Sugama06,Sugama08,Zonca08}, then included to higher orders~\cite{Xu08prl,Qiu09ppcf}. Finally, the effect of the flux surface shape was also included~\cite{Gao10pop}.
All analytical calculations of the Landau damping of GAMs have been performed so far by assuming a uniform plasma, i.e. by neglecting the effect of a gradient of the equilibrium density or temperature profiles.

More recently, the phase mixing, which was previously investigated for non-ionized fluids~\cite{Case60}, for Langmuir waves in plasmas~\cite{Sedlacek71}, and for shear-Alfv\'en waves in plasmas~\cite{Grad69,Hasegawa74,Chen74}, has also been investigated for GAMs analytically~\cite{Zonca08,Qiu11pst}. Contrarily to the Landau damping, the phase mixing can be studied also in fluid theory, whenever the local spectrum of oscillation of a wave - the {\it continuous spectrum} - varies in space (for the GAM in a tokamak, this happens due to the variation of the plasma temperature across the magnetic flux-surfaces).
In this case, the spatial shape of a perturbation will be deformed in time due to the effect of the different frequency of oscillation, occurring at different positions. When performing a Fourier transform in space, the effect of the phase mixing is that of shifting the peak of the perturbation in time, towards higher values of the wavenumber. Therefore, the amplitude of the perturbation in Fourier space calculated at the initial wavenumber, decays in time, giving rise to the {\it continuum damping}.
Gyrokinetic simulations of GAMs with realistic profiles of ASDEX Upgrade had already shown a peculiar behaviour at the edge, which was not been possible to explain with a description based on the Landau damping alone~\cite{Biancalani13EPS}. In particular the damping rate at the edge, where the temperature profile is steep, was measured to be much stronger than that predicted by the theory of Landau damping. More recently, the investigation of this strong damping observed in gyrokinetic simulations has started, showing a new mechanism based upon the combined action of the phase mixing and the Landau damping, and the first results have been shown in Ref.~\cite{Palermo16}.

In this work, we complete the level of understanding of this problem, by describing the details of the combined effect of the nonuniformities in velocity- and in real-space, causing the cascade in the wavenumber and the effective damping at high wavenumbers, with a dedicated initial-value code and diagnostics. In fact, as previously mentioned, the phase mixing, acting because of the nonuniformity in space, damps only each single mode in Fourier space by continuum damping, but in the whole real space the energy of the electric field is conserved, so the actual total damping is zero.
Therefore, although our analysis is linear, we can state that the values of the damping rate calculated as pure Landau damping and as pure phase mixing do not just {\it linearly} sum up.
In fact, the Landau damping of GAMs strongly increases with the radial wave number, due to the finite orbit width of the passing ions. Consequently the effect of the phase mixing is to greatly amplify the efficiency of the Landau damping.
This general effect, recently proved to be crucial (see Ref.~\cite{Palermo16}), is explored here in specific cases chosen for their practical experimental relevance, and quantitatively proved to be indispensabile for an interpretation of realistic GAM dynamics. In particular, the phase mixing and Landau damping are first studied separately and then the resulting combined effect is shown to form a cascade of the energy in wavenumber (analogously to that generated in a turbulent system) with a strong absorption at the small scales~\cite{Hasegawa74,Chen74,Zonca08}, which is reproduced here quantitatively with a dedicated initial value code.
The investigation of the Landau damping with analytical theory and numerical simulations with ORB5 was performed already in Ref.~\cite{Biancalani14nufu,Palermo16}, and recalled here for sake of completeness. The investigation of the phase mixing, on the other hand, is performed in this work with the help of an initial value code which solves the GAM dynamics by adopting the real part of the kinetic frequency, and neglects its imaginary part. This initial value code for the study of the phase mixing, is used to reproduce the evolution of the radial structure of the GAM perturbation, modified by the phase mixing alone.

In tokamaks, the combined effect of Landau damping and phase mixing can play a crucial role in the dynamics and in the decay of GAM at the tokamak edge, where the temperature gradients are very large. Depending on the confinement regime, low confinement (L-mode), improved (low) confinement (I-mode) and high confinement (H-mode), different parameter values are achieved in the experiments, in particular changing the relative strength of temperature and density gradients at the edge of the plasma.
The sequence of events in the transition phases between these regimes, particularly regarding the turbulence suppression, the development of the edge temperature and density gradients and the increase of the shear flow, is not completely established yet.
GAMs are considered to be potential key players in the dynamics of the transition to I- and to H-modes~\cite{Cziegler13}, and their absence can enhance the effects of the shear flow on the turbulence~\cite{Angelino06}.
GAMs are reported to be regularly observed in L-mode, and are also observed in I-mode, whereas they are not observed in H-mode. It will be shown that the proposed mechanism of GAM decay is consistent with the observed existence or non-existence of GAMs in the different confinement regimes~\cite{Conway11}.

Our investigation is carried out by means of analytical theory and numerical simulations. For the numerical simulations, we use the global particle-in-cell code ORB5~\cite{Jolliet07}, which now includes all extensions made in the NEMORB project~\cite{Bottino15jpp,Biancalani16cpc}. The ORB5 code uses a Lagrangian formulation based on the gyrokinetic Vlasov-Maxwell equations~\cite{Hahm88,Sugama00,Brizard07,Bottino11,Scott10}.
Due to the method of derivation of the GK Vlasov-Maxwell equations from a discretized Lagrangian, the symmetry properties of the starting Lagrangian are passed to the Vlasov-Maxwell equations, and the conservation theorems for the energy and momentum are automatically satisfied~\cite{Scott03}.
The code solves the full-f gyrokinetic Vlasov equation for ions and the drift-kinetic equation for electrons. Only linear collisionless electrostatic simulations are considered in this paper, with electrons treated as adiabatic, and with magnetic flux surfaces with circular concentric poloidal sections.

This paper is structured as follows. In Sec.~\ref{sec:benchmark}, the Landau damping in uniform plasmas is studied with ORB5 and compared with analytical theory. In Sec.~\ref{sec:phase-mixing}, the phase mixing signatures are identified and compared with results of numerical simulations, obtained for realistic values of tokamak plasmas. Sec.~\ref{sec:Landau-phasemixing} is devoted to a description of the combined action of the phase mixing and Landau damping. Finally, Sec.~\ref{sec:LIH-modes} discusses the application to predict the regimes where such damping overcomes the drive given by the turbulence, and the summary of the results.

\section{Landau damping in uniform plasmas}
\label{sec:benchmark}


In this Section, we recall the results about the Landau damping of GAMs in uniform plasmas, i.e. in the absence of gradients of the equilibrium profiles (i.e. the safety factor, density and temperature profiles are flat), shown in Ref.~\cite{Palermo16}. 
The GAM dynamics is independent on the radius to a good approximation, and the radial shape of the initial perturbation does not change in time, at least for the first few GAM oscillations (after which, global effects can occur).
For our regimes of interest, a linear estimation of the frequency and damping rate is given by the analytical theory where first-order ion finite-orbit-width effects are retained~\cite{Sugama06,Sugama08}.
These simulations serve as a verification of ORB5, extending the work previously done in Ref.~\cite{Biancalani14nufu}. In particular, we study the dependence of the damping rate on the radial wave-number $k_r$, for different values of the safety factor $q$.

We initialize a zonal perturbation of the radial component of the electric field, i.e. a perturbation independent of poloidal and toroidal angles, and depending only on the radius. In general in this paper, except where otherwise stated, we consider initial perturbations with a sinusoidal radial dependence, i.e. monochromatic in Fourier space. The perturbation evolves in time in a linear electrostatic collisionless simulation with adiabatic electrons.
For each simulation, we investigate the evolution in time of the radial electric field, and in particular the values frequency and damping rate at each radial location, and the radial structure at each time. This type of numerical experiment is known as a Rosenbluth-Hinton test~\cite{Rosenbluth98}.

Different simulations are investigated, with wavenumber chosen within the range 0.02 $\le k_r\rho_i \le$ 0.2, and safety factor within the range 1 $\le$ q $\le$ 3.5. 
The value of the plasma temperature is defined by $\rho^* = \rho_s/a = 1/160$. The ion Larmor radius is defined as $\rho_i =   \sqrt{2} \sqrt{T_{i0}/T_{e0}}\rho_s$, with $\rho_s = \sqrt{T_{e0}m_i}/(eB_0)$ and with $T_{i0}/T_{e0}$ being the ion/electron temperature calculated in the middle of the radial domain (in this Section, $T_e = T_i = T_{e0} = T_{i0}$). The magnetic field $B_0$ is calculated at the magnetic axis (r=0).
We choose a tokamak with an inverse aspect ratio $\epsilon = a/R_0$ = 0.1. We consider analytical magnetic equilibria with circular flux surfaces. In this limit, the flux surface coordinate $r = \sqrt{\psi/\psi_{edge}}$ is a good approximation of the radial coordinate (normalized to the minor radius $a$).

The reference simulation has a spatial grid of ($r \times \theta \times \phi$) = ($256 \times 64 \times 4$) and a time step of 100 $\Omega_i^{-1}$, with $\Omega_i$ being the ion cyclotron frequency. All the simulations have been performed with $10^8$ ion markers.
As a general remark, we find very good agreement between theory and simulations, consistently with Ref.~\cite{Biancalani14nufu}. In particular, our results with ORB5 are consistent with the analytical theory, showing that, at realistic values of q, the damping rate strongly increases with $k_r\rho_i$ (see Ref.~\cite{Sugama06,Sugama08,Palermo16})).

\section{Phase mixing in the presence of a radial nonuniformity}
\label{sec:phase-mixing}

In this Section, we investigate the effect of a radial nonuniformity of the equilibrium, in generating the phase mixing. Ronsebluth-Hinton tests similar to those described in the previous Section are performed, but with the main difference that the equilibrium radial profiles are not considered flat.
This introduces a radial dependence in the GAM dynamics, for example of the GAM local frequency: $\omega_G = \omega_G (r)$. In this case, the GAM dynamics in its simplest form (no Landau damping considered here) can be described by the vorticity equation~\cite{Zonca08,Qiu11pst}:
\begin{equation}\label{eq:vorticity}
\frac{\partial}{\partial r} \Big( \frac{\partial^2}{\partial t^2} - \omega^2_G (r)  \Big) \frac{\partial}{\partial r} \phi(r,t) = 0 
\end{equation} 
This equation has been shown to describe the phase mixing~\cite{Case60,Sedlacek71,Grad69,Hasegawa74,Chen74}, i.e. the cascade of energy from bigger to smaller spatial scales. We can recover the phase mixing by integrating Eq.~\ref{eq:vorticity} in space once, which yields the radial component of the electric field $E(r,t)=E_0 \exp(-i\omega_G(r) t)$, and approximating for simplicity the frequency profile as linear in space: $\omega_G(r)\simeq \omega_{G0} + \omega_G' \cdot (r-r_0)$.
Now it is sufficient to calculate the Fourier transform of the electric field in a radial region centered at $r_0$, and with half-width $\nu$, to obtain~\cite{Palermo16}:
\begin{equation}
E(k_r,t) = 2 E_0 \exp(-i \omega_{G0} t) \exp(-i k_r r_0) \sin((\omega_G'  t - k_r) \nu) / (\omega_G' t - k_r)
\end{equation}
which can be written, in the limit of radially localized perturbation, as:
\begin{equation}
E(k_r,t) = 2\pi E_0 \exp(-i\omega_{G0} t) \exp(-i k_r r_0) \delta(k_r - \omega_G' t)
\end{equation}
As a first signature of the phase mixing, we observe the cascade of energy in Fourier space, in the form of a linear dependence of the wavenumber on time:
\begin{equation}\label{eq:phase-mixing}
k_r \propto \omega_G' t
\end{equation}
As a second signature, we observe that the scalar potential dependence on time can be now estimated from $E(r,t)=i k_r(t) \phi(r,t)$, obtaining for its amplitude the characteristic decay which takes the name of {\it continuum damping}:
\begin{equation}
|\phi(r,t)| \propto (\omega_G' t)^{-1}
\end{equation}
Before proceeding further, we note that locally the phase-mixing does not dissipate the energy of the GAM, being the local total energy (ExB + thermodynamic) equal to $d\mathcal{E}(t) = \rho_m |v_E(t=0)|^2 /2 \, dr$, with $\rho_m$ being the mass density, and $v_E = c E/B_0$ being the ExB drift. The investigation of the global properties of GAMs, including the radial propagation of energy, are out of the scope of this paper, and will be discussed separately.

\begin{figure}[b!]
\begin{center}
\includegraphics[width=0.49\textwidth]{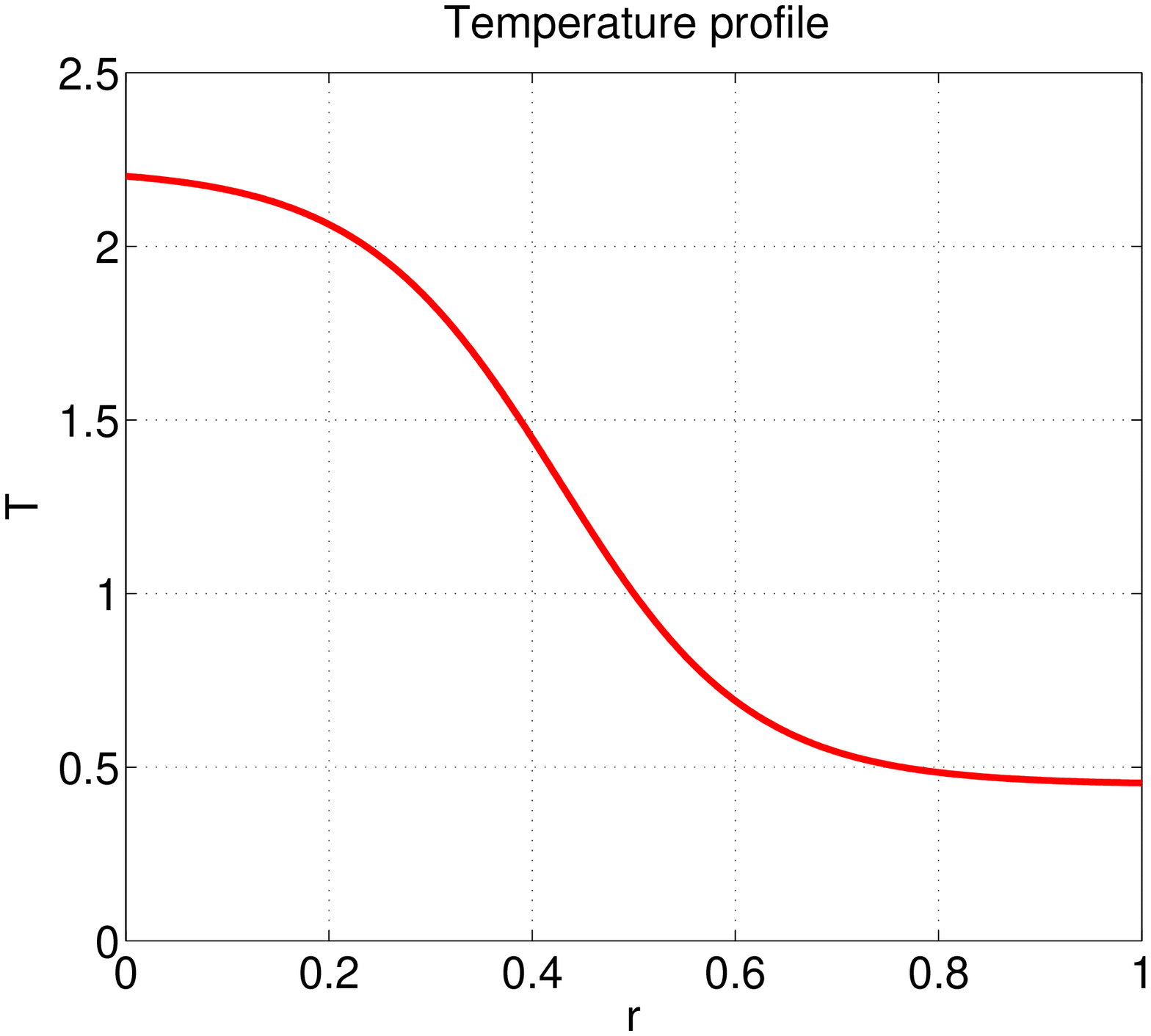}
\includegraphics[width=0.49\textwidth]{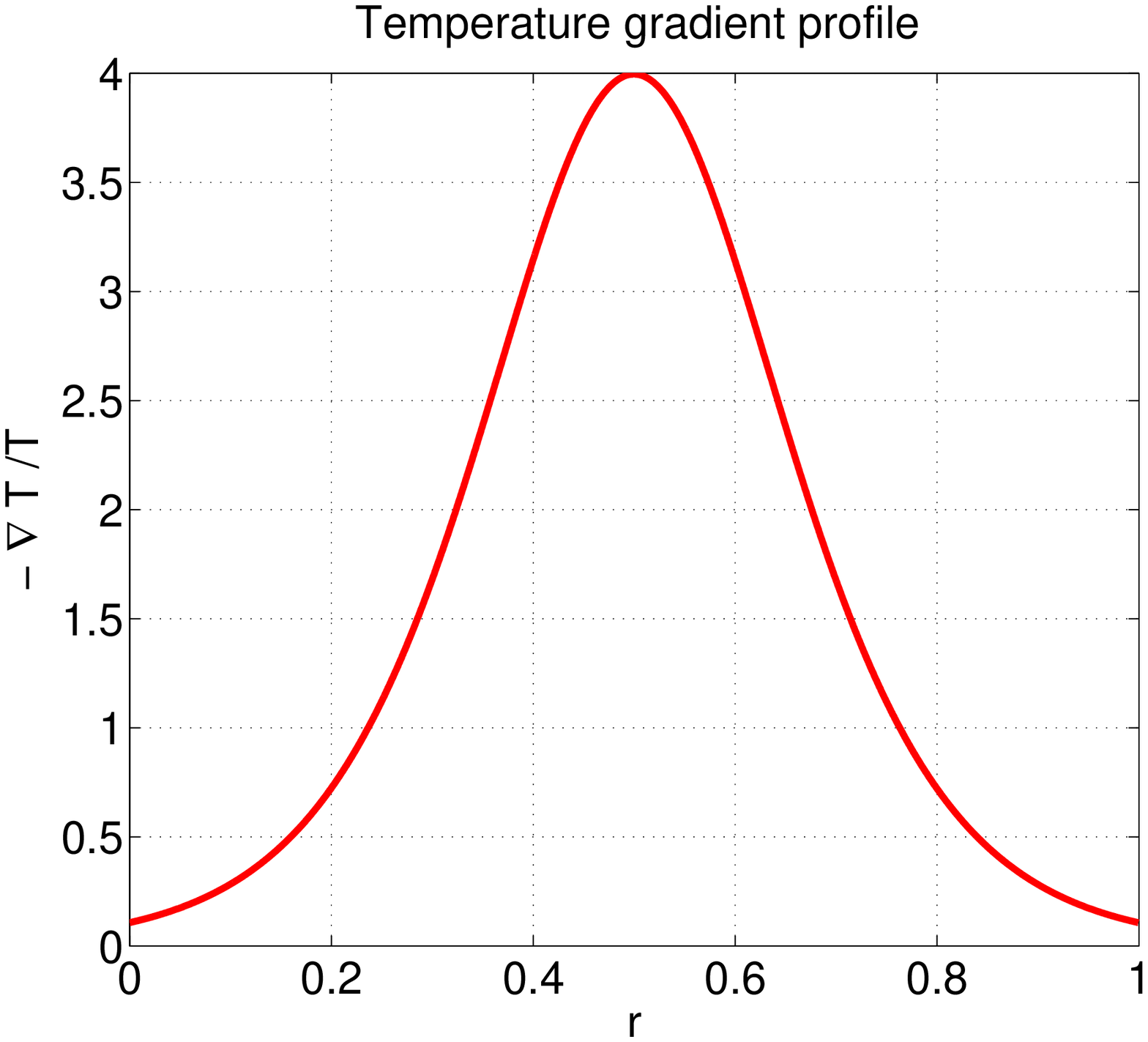}
\caption{\label{fig:T-prof} Temperature profile normalized at r=0.5 (left) and temperature gradient (right), for a simulation with $k_T a = 4$ at r=0.5.}
\end{center}
\end{figure}

We now want to investigate the signatures of the phase mixing in the results of the numerical simulations in an equilibrium obtained from realistic values of parameters. 
To this purpose, we have selected a group of parameters defining the equilibrium of ASDEX Upgrade for the shot number 20787~\cite{Conway08}, at the radial location where the GAM is detected (r $\simeq$ 0.9 a), consistently with Ref.~\cite{Biancalani13EPS,Palermo16}. These parameters will be used for the simulations described in the rest of the paper.
The magnetic equilibrium is characterized by a magnetic field on axis of $B_0=2$ T, the major and minor radii of $R_0=1.65$ m and $a=0.5$ m, and a safety factor of $q=3$.
The plasma has a local deuterium and electron temperatures of $T_i=T_e=170$ eV. We want to investigate the results of several simulations, with different temperature profiles with temperature gradients in the range $k_T a = a \nabla T/T =1 - 16$ at r=0.5 (density and safety factor profiles are kept flat for simplicity). The GAM characteristic radial size is estimated as $\lambda_{GAM} \simeq 10$ cm. Starting from these values we obtain the following quantities: an ion cyclotron frequency $\Omega_i = 0.96 \cdot 10^8$ rad/s, an ion thermal velocity $v_{thi} = \sqrt{2 T_i/m_i} = 0.9 \cdot 10^5$ m/s and a Larmor radius $\rho_i = 1.33 \cdot 10^{-3}$ m (with $\rho^* = 1/530$).
We initialize an electric field with sinusoidal profile with $k_r a = 10 \pi$, equivalent to $ k_r\rho_i = 0.084 $. 
We note that the GAM frequency depends strongly on the plasma temperature (for the order of magnitude we have $\omega_{GAM} \sim c_s/R \propto \sqrt{T}$, whereas $\omega_{GAM}$ does not depend on $n$ and depends weakly on $q$ in the considered regime of parameters), therefore we start by considering a simplified problem where the density and safety factor profiles are considered flat and the temperature profile has a radial dependence described as (see Fig.~\ref{fig:T-prof}):
\begin{equation}
\frac{T(r)}{T(r_0)} = \exp\Big( \Delta \, k_T \tanh \Big(\frac{r-r_0}{\Delta} \Big) \Big)
\end{equation}
where $r_0=0.5$ is the reference position in the simulation where the peak of the temperature gradient is located, and where the measurements are performed, $k_T = d \ln(T) /dr$ is the value of the temperture gradient, and $\Delta=0.2$ is the radial size of the temperature gradient peak (in this paper, electron and ion temperatures profiles are always considered equal).

\begin{figure}[b!]
\begin{center}
\includegraphics[width=0.49\textwidth]{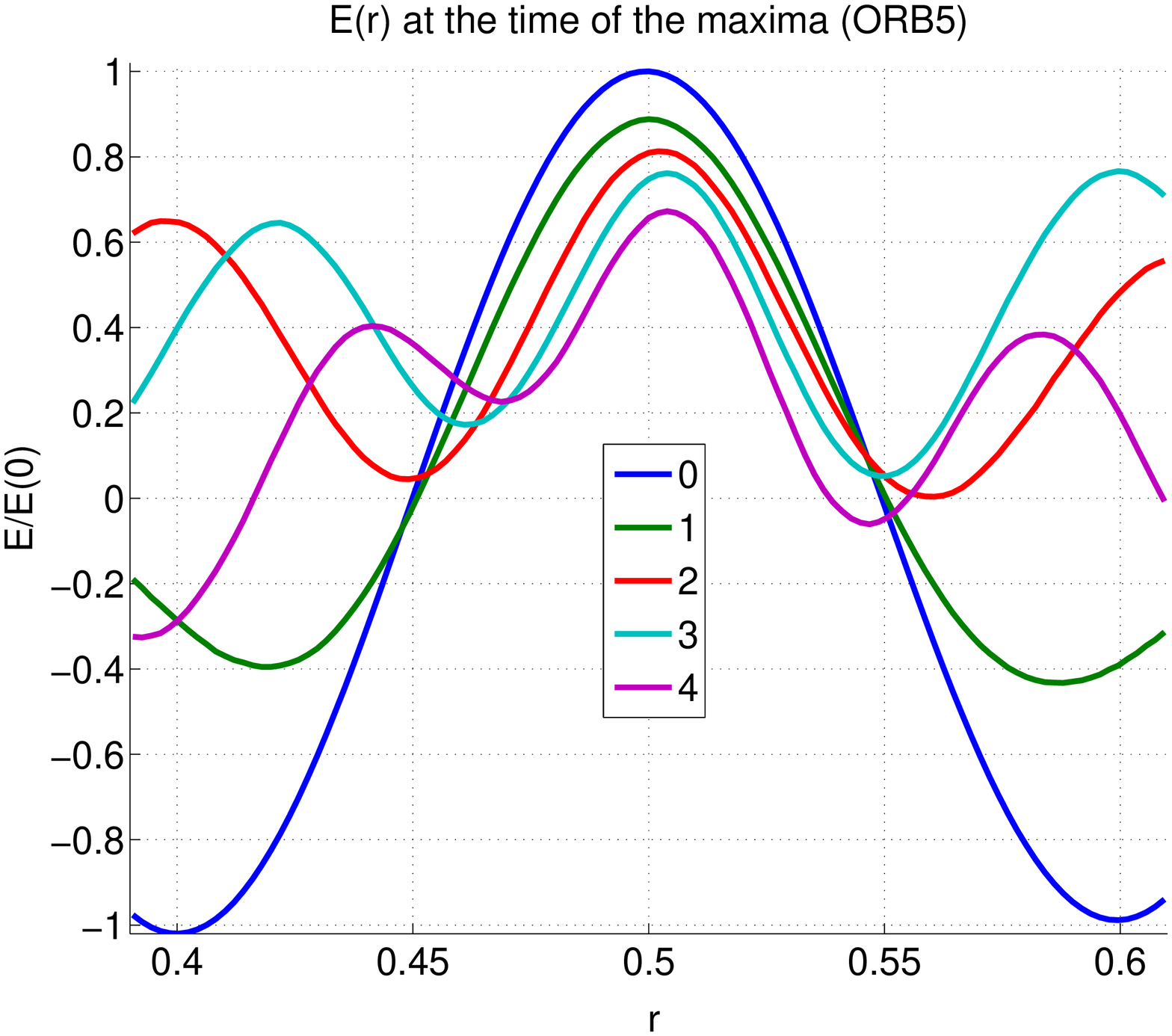}
\includegraphics[width=0.49\textwidth]{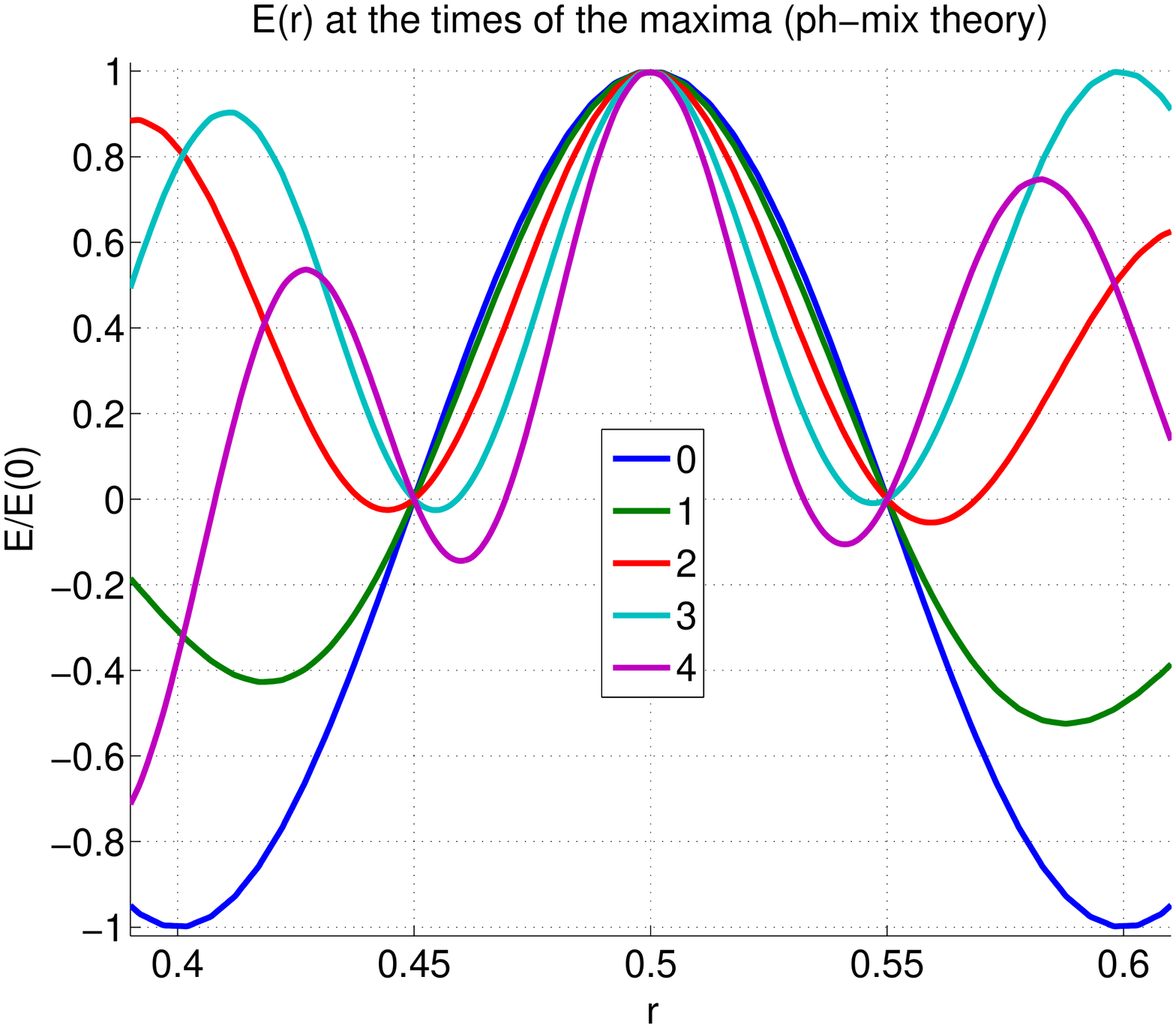}
\caption{\label{fig:phase-mixing-shape} On the left, zoom of the electric field profile measured with ORB5 at 5 different times when the maximum at r=0.5 is reached, for a simulation with $k_T a = 4$ at r=0.5. On the right, the same simulation performed with the phase-mixing initial value code. The blue line, labelled as ``0'', is the initial value of the simulations.}
\end{center}
\end{figure}

The results of these simulations are investigated, with a particular attention in this paper to the evolution of the radial shape of the perturbation (see Fig.~\ref{fig:phase-mixing-shape}). In order to isolate the physics of the phase mixing, we repeat the simulation with a dedicated initial value code, which evolves the electric field at each position with its local frequency:
\begin{equation}
E_{ph-mix}(r,t) = E(r,0)\exp(i \, \omega_G(r)t)
\end{equation}
where the frequency is given at each position by the analytical theory, and depends on the local temperature value. No Landau damping is considered in this code. This initial value code gives the evolution in time of the radial structure of the perturbation, due to the pure phase-mixing. Effects of the Landau damping in modifying the radial structure are neglected in this code. The comparison of the results of ORB5 and the phase-mixing code are shown in Fig.~\ref{fig:phase-mixing-shape}.
The first feature of the radial structure evolution, observed in both codes, is the increase of the wavenumber in time. This can be observed to occur in both ORB5 and in the phase-mixing code, with similar values. We conclude that the wavenumber calculated in gyrokinetic simulation is approximated to a good extent by the phase-mixing code. As a second remark, we note that the results of ORB5 show a damping in time at r=0.5, whereas the phase-mixing code does not show it, for construction. This damping will be investigated in  the next Section.
As a third remark, we note that the results of ORB5 show a slight displacement in time of the position of the peak near r=0.5. This second-order global effect due to the radial GAM propagation, which is not present in the phase-mixing code, will be investigated in a separate paper.

\begin{figure}[t!]
\begin{center}
\includegraphics[width=0.47\textwidth]{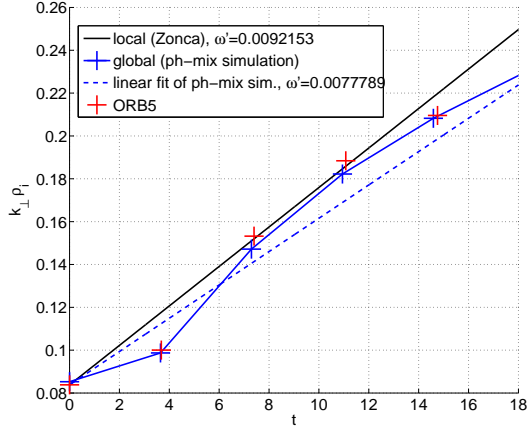}
\caption{\label{fig:phase-mixing-khat_t} Evolution in time of the wavenumber measured with ORB5 (red crosses), predicted analytically in the local limit (black line) and with initial-value code with first order global effects (blue crosses). The linear fit of the effective wavenumber cascade is also shown with a blue dashed line.}
\end{center}
\end{figure}

As time goes on, we measure the wavenumbers at the center of the radial domain, for both results of ORB5 and of the phase-mixing code, and plot them in time. The result can be seen in Fig.~\ref{fig:phase-mixing-khat_t}, where the value of the analytical prediction given by the local approximation, Eq.~\ref{eq:phase-mixing}, is also shown as a black line.
One can see that the wavenumber measured in the numerical simulations is found to grow more slowly than the local analytical prediction on average. This is due to the local approximation, which does not consider the finite size of the temperature gradient peak. In fact the GAM perturbation experiences a  lower temperature gradient away from the reference position r=0.5, and therefore the global effect is an average of the temperature gradient profile, rather than the value measured at r=0.5 only. In Fig.~\ref{fig:phase-mixing-khat_t}, the analytical prediction taking into account also this average to the first order is shown.
One can see that the behaviour of the wavenumber measured with the numerical simulations is described for a longer time by the analytical prediction where the global effects are considered to the first order, and eventually higher order effects should be also taken into account. In conclusion, we observe that the phase mixing affects the numerical simulation and we can now extend this result to the combined effect of Landau damping and phase mixing.

\section{Combined effect of phase mixing and Landau damping}
\label{sec:Landau-phasemixing}

We are now ready to investigate the combined effects of the phase mixing and Landau damping, which have been studied separately in the previous Sections. The same simulations as in Sec.~\ref{sec:phase-mixing} are considered, where the realistic conditions at the tokamak edge of AUG shot $\#$ 20787 taken in Ref.~\cite{Conway08}, are reproduced at the center of the radial domain of the simulations. As shown in Fig.~\ref{fig:phase-mixing-khat_t}, the radial shape of the electric field perturbation evolves in time with higher and higher wavenumber, consistently with Ref.~\cite{Zonca08,Qiu11pst}. In the same Figure, we can observe also that the value of the electric field at r=0.5 is lower and lower at each oscillation. In this Section, we want to investigate this damping with a proper theoretical model~\cite{Palermo16}.

\begin{figure}[b!]
\begin{center}
\includegraphics[width=0.47\textwidth]{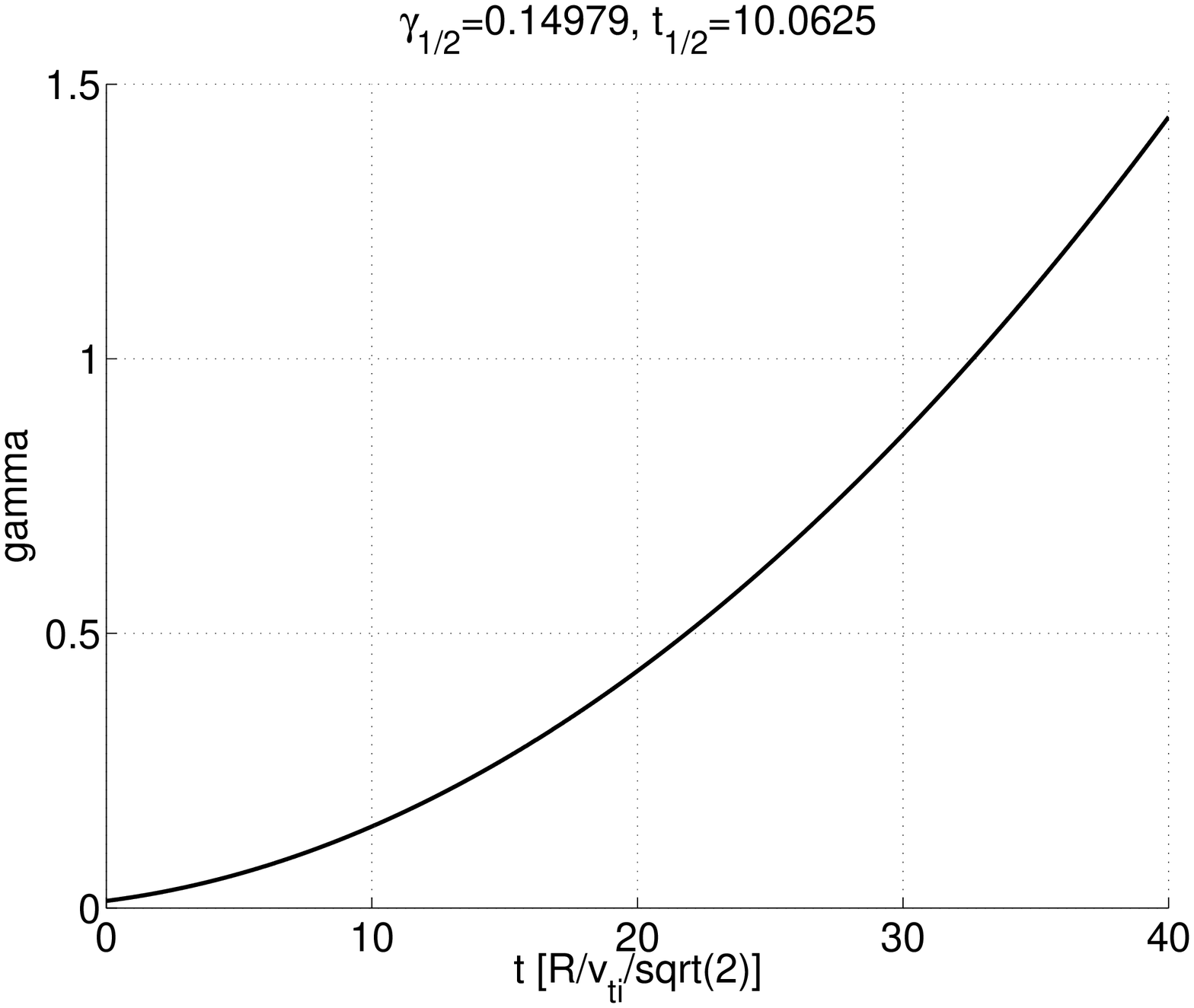}
\includegraphics[width=0.47\textwidth]{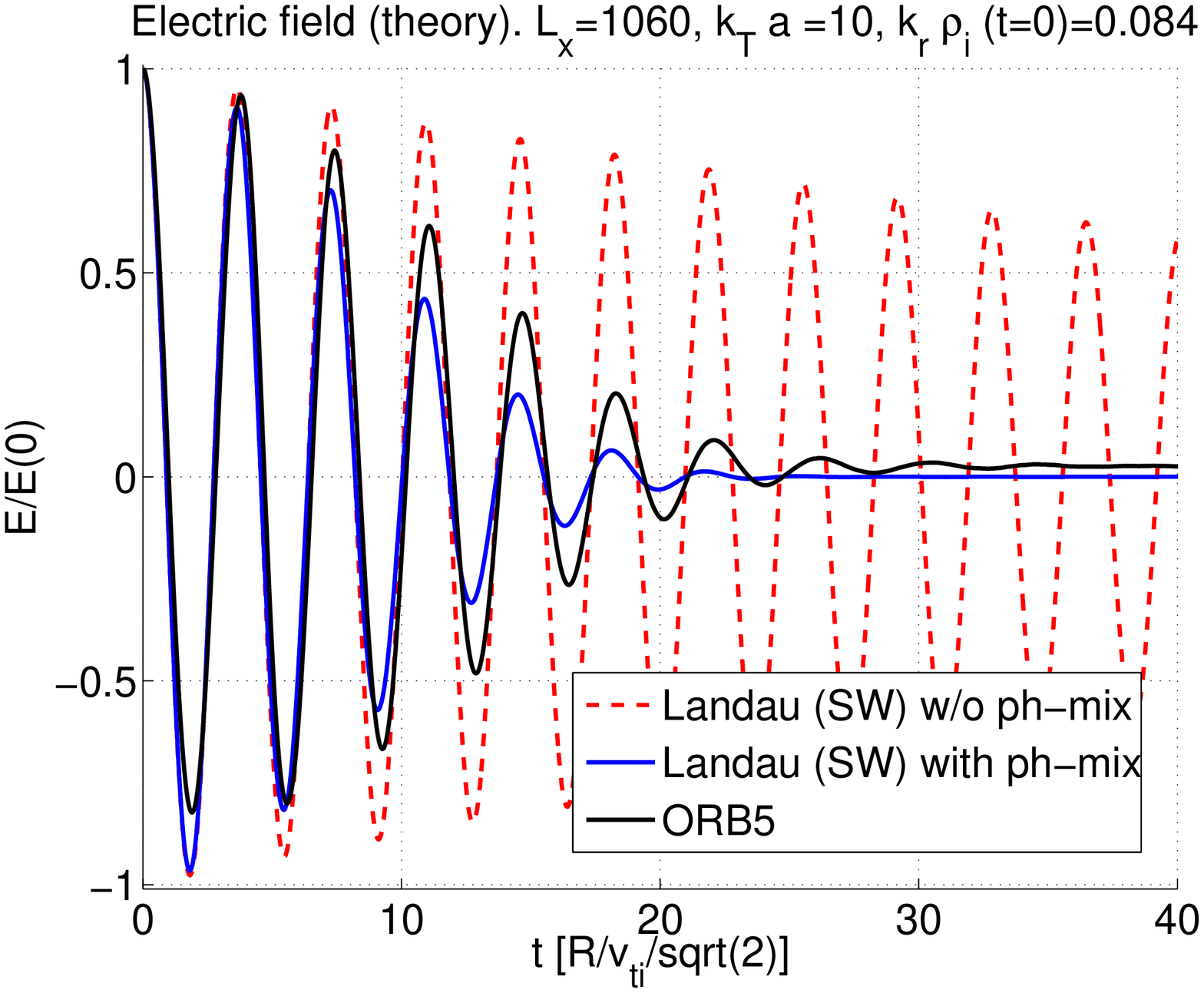}
\caption{\label{fig:landau-phmix-2D} On the left, the damping rate as a function of time for a simulation with $k_T a = 10$. On the right, electric field measured in the middle of the radial domain for the same simulation. The black line is measured in the gyrokinetic simulation, and it is compared with the red dashed line, given by the theory of the Landau damping alone, and with the blue line, given by the PL mechanism.}
\end{center}
\end{figure}

The values of q and $k_r \rho_i$ allow us to treat the Landau damping with the analytical theory with first-order FOW effects (i.e. with a quadratic dependence on $k_r \rho_i$). We can then write the damping rate due to the combined action of phase mixing and Landau damping (PL) as a function of time:
\begin{equation}\label{eq:gamma_PL-1}
\gamma_{PL}(k_T,t) = f(v_{Ti},q,\tau_e) + (k_{r0}+ \omega_G' t)^2 g(v_{Ti},q,\tau_e) 
\end{equation}
where f and g are the cofficients given in Ref.~\cite{Sugama06,Sugama08}. Consequently, the evolution of the electric field envelop is described at each time by:
\begin{equation}\label{eq:gamma_PL-2}
\frac{1}{E} \frac{\partial E}{\partial t} = \gamma_{PL}(k_T,t)
\end{equation}
Equation~\ref{eq:gamma_PL-2} is solved with an initial-value code in order to calculate theoretically the evolution in time of the GAM electric field envelop. The frequency is also known from Ref.~\cite{Sugama06,Sugama08}, and does not change in time to a good approximation. The comparison of the analytical prediction with the result of ORB5 is shown in Fig.~\ref{fig:landau-phmix-2D}. 
We note that the GAM decay is described to a good extent by the PL model, whereas the Landau damping alone greatly overestimates the GAM amplitude at large times. This is due to the fact that the damping rate, which is the Landau damping corresponding to the initial wavenumber at t=0, grows quadratically in time due to the evolution of the wavenumber (see Fig.~\ref{fig:landau-phmix-2D}). We also note that some discrepancy is present, due to global effects, such as the initial formation of the GAM eigenmode during the first oscillation, and the radial displacement of the GAM perturbation at later times. The analysis of these corrections due to radial propagation will be done in a separate paper.

\begin{figure}[t!]
\begin{center}
\includegraphics[width=0.5\textwidth]{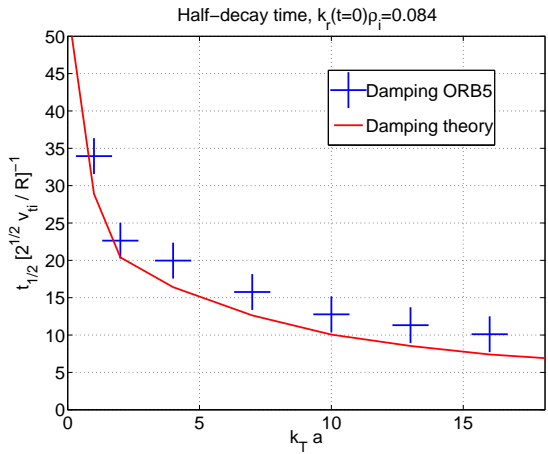}
\caption{\label{fig:thalf_kT} Half-decay time for simulations with different temperature gradients. The results of numerical simulations are shown with blue crosses, and the theory with a red continuous line.}
\end{center}
\end{figure}

We now investigate the dependence of the GAM decay on the value of the temperature gradient, with both gyrokinetic simulations and the analytical theory of the PL mechanism. In particular, the evolution in time of the electric field is measured at r=0.5 in gyrokinetic simulations with different values of $k_T$ ranging from $2 \le k_T a \le 16$. The time of half-decay is measured for each simulation and shown in Fig.~\ref{fig:thalf_kT}, where it is compared with the analytical model of the PL mechanism. 
We observe that the decay time strongly decreases with the increase of the temperature gradient. 
The theory for the evolution of the envelop of the electric field, discussed in Sec.~\ref{sec:Landau-phasemixing}, describes well the strong damping with respect to the case $k_T a$ = 0 in which only Landau damping acts. We note that for large values of $k_T a$, the half-decay time in the simulations is slightly larger than the damping expected by the theory, which is due to the radial propagation not investigated in this paper.

\section{Summary and conclusions.}
\label{sec:LIH-modes}

Understanding the linear damping mechanisms of GAMs is a crucial step for reaching a complete understanding of their dynamics, and in particular of their nonlinear interaction with turbulence in tokamaks. 
Landau damping has been historically recognized as the main linear damping mechanism of GAMs, due to their low frequency (for instance with respect to other higher frequency, less damped modes, like incompressible shear-Alfv\'en modes). Landau damping has been shown to depend on the radial shape of GAMs, in particular increasing with the local radial wavenumber. Moreover, the phase mixing has been predicted to affect GAMs, by means of analytical theory~\cite{Zonca08,Qiu11pst}. The main signature of phase mixing has been shown to be a cascade of the GAM energy from lower to higher values of the wavenumber. In previous works~\cite{Biancalani13EPS,Palermo16}, the results of gyrokinetic simulations in the presence of a temperature gradient have been investigated, and in Ref.~\cite{Palermo16} the combined effect of the phase mixing and the Landau damping has been shown to generate a novel damping mechanism, responsible for a very efficient absorbption of the GAM energy by the bulk plasma~\cite{Palermo16}. This 
phase mixing / Landau 
damping (PL) mechanism, has been shown to increase up to an order of magnitude the effective damping of GAMs, for realistic tokamak conditions. 

In this work, we have investigated the details of the radial structure evolution in time of the GAM electric field in the presence of a temperature gradient. The cascade of the wavenumber in time has been investigated with gyrokinetic simulations and with a specific initial value code dedicated to the investigation of the phase-mixing. The radial structure of gyrokinetic simulations obtained with ORB5 has been shown to be approximated to a good extent by the one obtained with the initial value code. Second order effects, like the radial displacement of the GAM peak in time, have been neglected at this stage and have been left for a future work. Finally, consistently with Ref.~\cite{Palermo16}, the combined effect of the phase-mixing and Landau damping has been investigated, and shown to decrease the half-decay time up to one order of magnitude, for realistic tokamak parameters, and using new simulations where the temperature gradient is peaked at the center of the radial domain.

From these results, we deduce that the PL mechanism can play an important role in the suppression of GAM oscillations in the regions characterized by a strongly nonuniform temperature profile such as in the H and I modes. We note that simulations performed with a density gradient different from zero have given results very close to the simulations performed with a flat density profile.
This is in agreement with Eq.~\ref{eq:gamma_PL-1} that does not depends on the density gradient. Moreover, we note that $\gamma_{PL}$ in Eq.~\ref{eq:gamma_PL-1} depends also on q. However, the gradient of q has a weak influence on the phase mixing and consequently on $\gamma_{PL}$. This aspect has been verified by numerical simulations showing that the main parameter in the PL mechanism is the temperature gradient.

In order to investigate and to quantify the importance of the PL damping mechanism on these regions, we compare it with the characteristic drive rate $\gamma_{RD}$ given by the nonlinear coupling with the ITG mode in Ref.~\cite{Zonca08}, and discussed in Ref.~\cite{Palermo16}. 
The characteristic drive time can be calculated as $t_{RD} \simeq 1/\gamma_{RD}$, and should be compared with the PL decay time. If $t_{PL} < t_{RD}$, this means that the PL damping rate exceeds the energy transfer rate from the ITG turbulence to the GAM, and as a consequence we may expect no observation of GAMs. The drive time is found to be $t_{RD} < t_s$ for the L-mode, $t_{RD} \sim t_s$ for the I-mode, and $t_{RD} \sim 10 \, t_s$ for the H-mode, with $t_s = 2^{-1/2}R/v_{ti}$ being the sound time unit. Therefore, we observe that the PL model explains the observation of GAMs for L and I modes, but a desappearance is predicted for the H-mode, for $k_T a > 10$ (see Ref.~\cite{Palermo16} for the detailed calculation). We note that this analysis of orders of magnitude gives results which are  consistent with experimental results that show the existence of GAM in I-mode regime in spite of the strong temperature gradient comparable to that of H-mode~\cite{Cziegler13,Conway11}. These results are also in agreement 
with the dynamics of GAM observed at the I-H transition~\cite{Cziegler13}.

As a future work, a more detailed description of a particular experimental shot will be done by using the micro-turbulence features observed in the experiment of AUG and comparing with linear and nonlinear gyrokinetic simulations where global effects are also investigated.

\begin{appendices}
 
\section{Numerical parameters and power balance.}

The numerical simulations described in this paper, have been performed with the global gyrokinetic particle-in-cell code ORB5, by solving electrostatic gyro-kinetic equations for the ions and treating the electrons as adiabatic. Only linear collisionless simulations have been performed. A typical simulations has a spatial grid of $(r, \theta, \phi)$ = 1024 x 64 x 4 and a time step of 50 $\Omega_i^{-1}$. The number of ion markers has been chosen as $N_i= 10^8$. A set of poloidal harmonics going from m=-10 to m=10 has been selected by means a filter in Fourier space in the poloidal direction. Dirichlet boundary conditions on the potentials have been imposed at the outer boundary, and Neumann boundary conditions at the inner boundary.

In a collisionless gyrokinetic simulation, the GAM amplitude is damped in time due to the combined effect of phase mixing and Landau damping, where the phase mixing creates a cascade in the radial wavenumber and the Landau damping effectively damps the higher wave numbers~\cite{Palermo16}. Ultimately, the energy of the GAM goes into the bulk ion microscopic kinetic energy due to the Landau damping. In this Section, this energy channel among the global GAM mode and the microscopic ion energy is investigated.

\begin{figure}[h!]
\begin{center}
\includegraphics[width=0.5\textwidth]{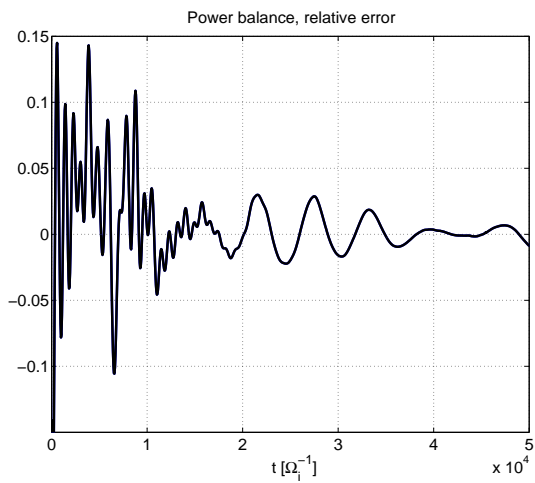}
\caption{\label{fig:loc-GAM-RH_run-check} The relative error of the power balance, for the simulation with $k_T a=10$, in a time zoom corresponding to about 10 GAM oscillations.}
\end{center}
\end{figure}

In this model the rate of change of the particle kinetic energy must be equal to the power transfer from the particles to the field (power balance) (see Ref.~\cite{Bottino15jpp}).
The power balance not only gives an indication of the quality of the simulations, but also
provides a measure of the instantaneous growth (decay) rate.
Defining $dE_k/dt$ the rate of variation of kinetic energy and $dE_f/dt$ the power transfer, the relative error of the power balance is:
\begin{equation}
\delta E_{rel} = \frac{ dE_{f}/dt - dE_k/dt}{dE_{f,max}/dt}
\end{equation}
where $dE_{f,max}/dt$ is the maximum value of $dE_{f}/dt$ in the time interval of interest. In Fig.~\ref{fig:loc-GAM-RH_run-check}, the relative error of two simulations is shown, namely one with flat temperature profile, and one with $k_T a=10$. For this specific runs, that falls within 3\% at the time of half decay.

\end{appendices}

\section*{Acknowledgements}
Interesting and useful discussions with G. Conway, E. Poli, P. Manz, Z. Qiu, B. Scott and ASDEX Upgrade team are kindly acknowledged.  This work has been done in the framework of the European Enabling Research Project on ``Verification and development of new algorithms for gyrokinetic codes``, WP15-ER-01/IPP-01, and the European Enabling Research Project on ``Micro-turbulence properties in the core of tokamak plasmas: close comparison between experimental observations and theoretical predictions'', WP15-ER-01/IPP-02. Simulations were performed on the IFERC-CSC Helios supercomputer within the framework of the VERIGYRO and the ORBFAST project.

\end{document}